\documentclass[preprint,aps,showpacs,nofootinbib,preprintnumbers,amsmath,amssymb]{revtex4-1}
\usepackage{}
\usepackage{epsfig}
\usepackage{subfigure}
\usepackage{dcolumn}
\usepackage{bm}
\usepackage[usenames ,dvipsnames]{xcolor}
\usepackage{slashed}
\usepackage{graphicx,color}

\begin{document}
\title{Branching fractions of $B_{(c)}$ decays involving $J/\psi$ and $X(3872)$}

\author{Y.K. Hsiao and C.Q. Geng}
\affiliation{
Chongqing University of Posts \& Telecommunications, Chongqing, 400065, China\\
Physics Division, National Center for Theoretical Sciences, Hsinchu, Taiwan 300\\
Department of Physics, National Tsing Hua University, Hsinchu, Taiwan 300
}
\date{\today}

\begin{abstract}
We study  two-body $B_{(c)}\to M_c(\pi,K)$ and semileptonic $B_{c}\to M_c\ell^-\bar \nu_\ell$ decays
with $M_c=(J/\psi,X_c^0)$, where $X_c^0\equiv X^0(3872)$ is regarded as the tetraquark state  of 
$c\bar cu\bar u(d\bar d)$. With the decay constant $f_{X_c^0}=(234\pm 52)$~MeV determined from the data,
we predict that ${\cal B}(B^-\to X_c^0\pi^-)=(11.5\pm 5.7)\times 10^{-6}$, 
${\cal B}(\bar B^0\to X_c^0\bar K^0)=(2.1\pm 1.0)\times 10^{-4}$, and
${\cal B}(\bar B^0_s\to X_c^0\bar K^0)=(11.4\pm 5.6)\times 10^{-6}$. 
With the form factors in QCD models, we calculate that
${\cal B}(B_c^-\to X_c^0\pi^-,X_c^0 K^-)=(6.0\pm 2.6)\times 10^{-5}$ and $(4.7\pm 2.0)\times 10^{-6}$, and
${\cal B}(B_c^-\to J/\psi \mu^-\bar \nu_\mu, X_c^0 \mu^-\bar \nu_\mu )=(2.3\pm0.6)\times 10^{-2}$ and
$(1.35\pm 0.18)\times 10^{-3}$, respectively, and extract the ratio of the fragmentation fractions to be 
$f_c/f_u=(6.4\pm 1.9)\times 10^{-3}$.
\end{abstract}

\maketitle
\section{introduction}
Through the $b\to c\bar c d(s)$ transition at quark level,
the $B$ decays are able to produce the $c\bar c$ bound states like $J/\psi$;
particularly, the hidden charm tetraquarks to consist of $c\bar c q\bar q'$, such as
$X^0(3872)$, $Y(4140)$, and $Z_c^+(4430)$, known as the $XYZ$ states~\cite{Chen:2016qju}.
For example, we have~\cite{pdg,BtoXcK_data}
\begin{eqnarray}\label{data1}
{\cal B}(B^-\to J/\psi K^-)&=&(1.026\pm 0.031)\times 10^{-3}\,,\nonumber\\
{\cal B}(B^-\to X_c^0 K^-)&=&(2.3\pm 0.9)\times 10^{-4}\,,
\end{eqnarray}
where $X_c^0\equiv X^0(3872)$ is composed of $c\bar c u\bar u(d\bar d)$,
measured to have the quantum numbers $J^{PC}=1^{++}$. On the other hand,
the $B^-_c$ decays from  the $b\to c\bar u d(s)$ transition
can also be the relevant production mechanism for
the $c\bar c$ and $c\bar c q\bar q'$ bound states. However,
the current measurements have been done only for the ratios,
given by~\cite{Aaij:2012dd,Aaij:2013vcx}
\begin{eqnarray}\label{data2}
{\cal R}_{c/u}&\equiv&\frac{f_c{\cal B}(B^-_c\to J/\psi \pi^-)}{f_u{\cal B}(B^-\to J/\psi K^-)}
=(0.68\pm 0.12)\%\,,\nonumber\\
{\cal R}_{K/\pi}&\equiv&\frac{{\cal B}(B^-_c\to J/\psi K^-)}{{\cal B}(B^-_c\to J/\psi \pi^-)}
=0.069\pm 0.020\,,\nonumber\\
{\cal R}_{\pi/\mu\bar \nu_\mu}&\equiv&
\frac{{\cal B}(B^-_c\to J/\psi \pi^-)}{{\cal B}(B^-_c\to J/\psi \mu^-\bar \nu_\mu)}
=(4.69\pm 0.54)\%\,,
\end{eqnarray}
where $f_{c,u}$ are the fragmentation fractions defined by $f_i\equiv {\cal B}(b\to B_i)$.
In addition, none of the $XYZ$ states has been observed in the $B_c$ decays yet.

 From Figs.~\ref{fig1}a and ~\ref{fig1}d,
the $B\to M_c M$ decays proceed with the $B\to M$ transition, which is followed by
the recoiled $M_c=(J/\psi,X_c^0)$ with $J^{PC}=(1^{--,++})$, respectively,
presented as the matrix elements of $\langle M_c|\bar c\gamma_\mu(1-\gamma_5) c|0\rangle$.
Unlike $J/\psi$ as the genuine $c\bar c$ bound state,
while the matrix element for the tetraquark production is in fact not computable,
$X_c^0$ is often taken as the charmonium state in the
QCD models~\cite{Meng:2005er,Liu:2007uj,Wang:2007sxa}.
In this study, we will 
extract $\langle X_c^0|\bar c\gamma_\mu(1-\gamma_5) c|0\rangle$
from the data of ${\cal B}(B^-\to X_c^0 K^-)$ in Eq.~(\ref{data1}) to examine
the decays of $B^-\to X_c^0 (\pi^-,K^-)$, $\bar B^0\to X_c^0 (\pi^-,K^-)$,
and $\bar B^0_s\to X_c^0 K^-$, of which the extraction allows $X_c^0$ to be 
the tetraquark state.
On the other hand, to calculate
the $B_c^-\to (J/\psi,X_c^0)M$ decays in Figs.~\ref{fig1}b and \ref{fig1}e
and the semileptonic $B_c^-\to (J/\psi,X_c^0) \ell\bar \nu_\ell$ decays in Figs.~\ref{fig1}c and \ref{fig1}f,
we use the $B_c\to M_c$ transition matrix elements from the QCD calculations.

\section{Formalism}
\begin{figure}[t!]
\centering
\includegraphics[width=2in]{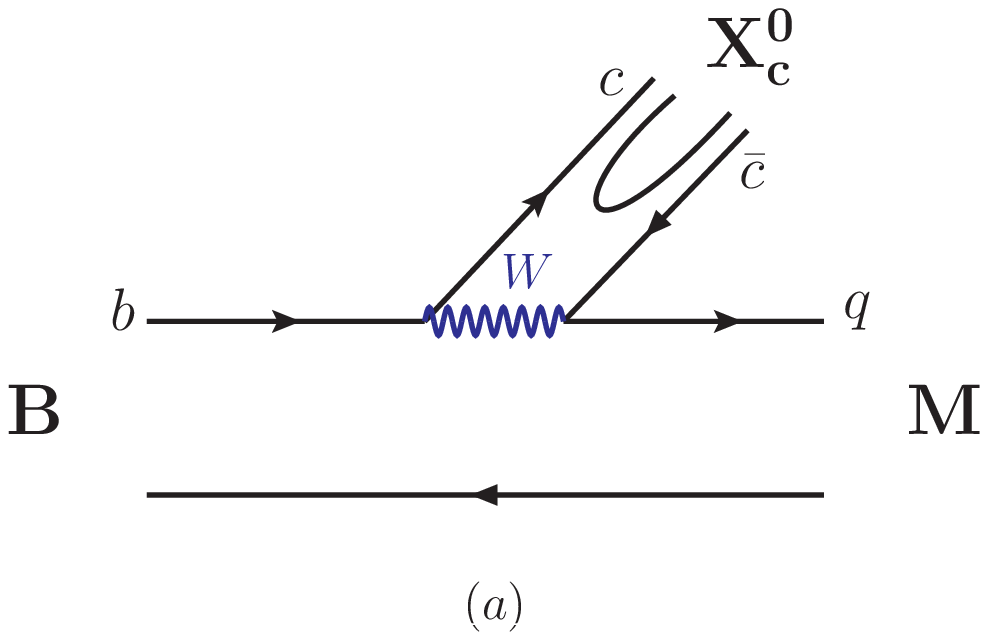}
\includegraphics[width=2in]{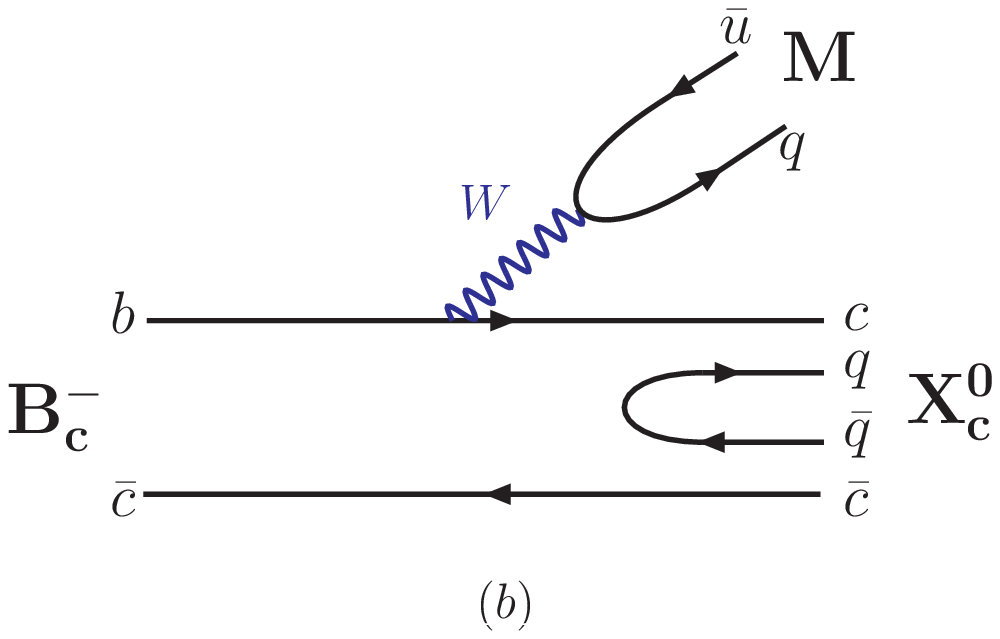}
\includegraphics[width=2in]{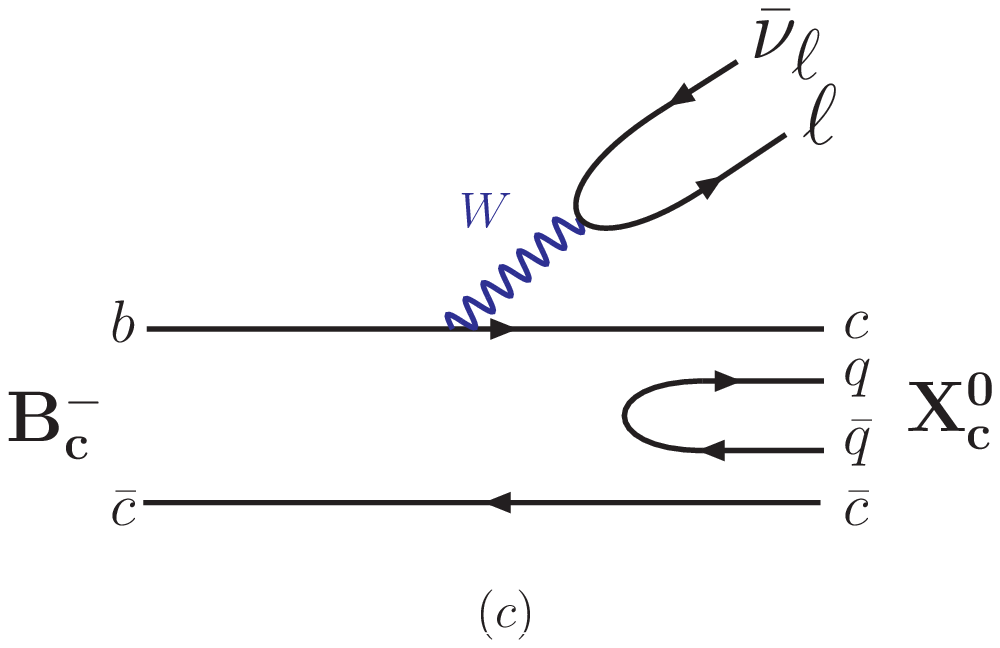}
\includegraphics[width=2in]{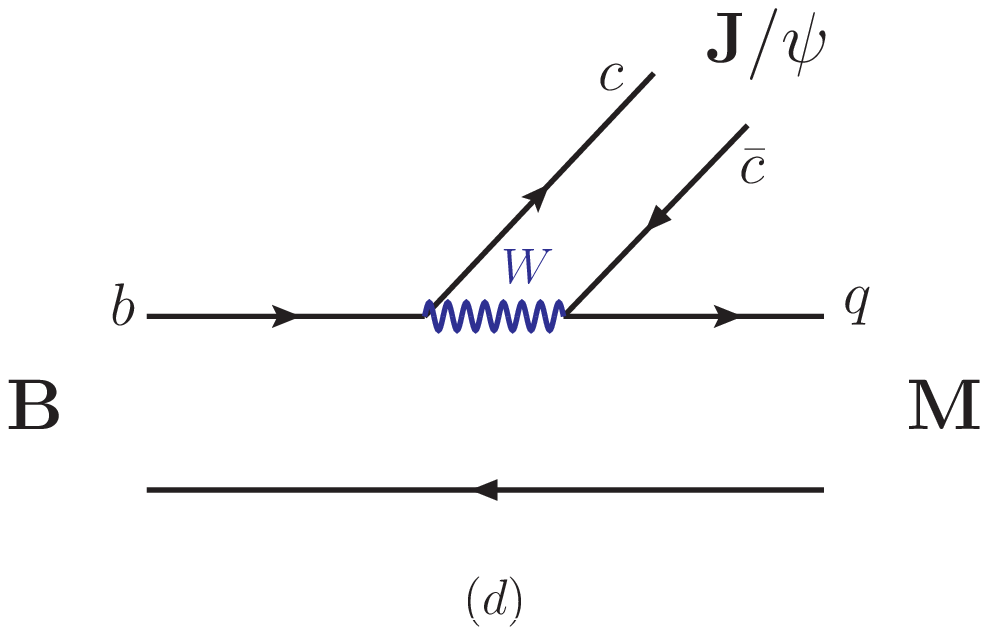}
\includegraphics[width=2in]{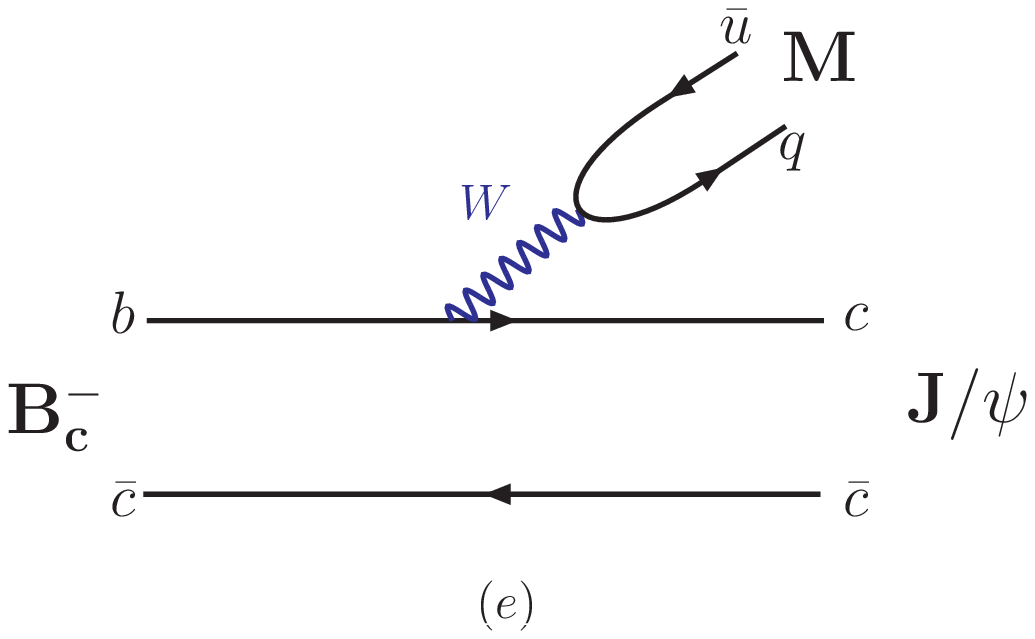}
\includegraphics[width=2in]{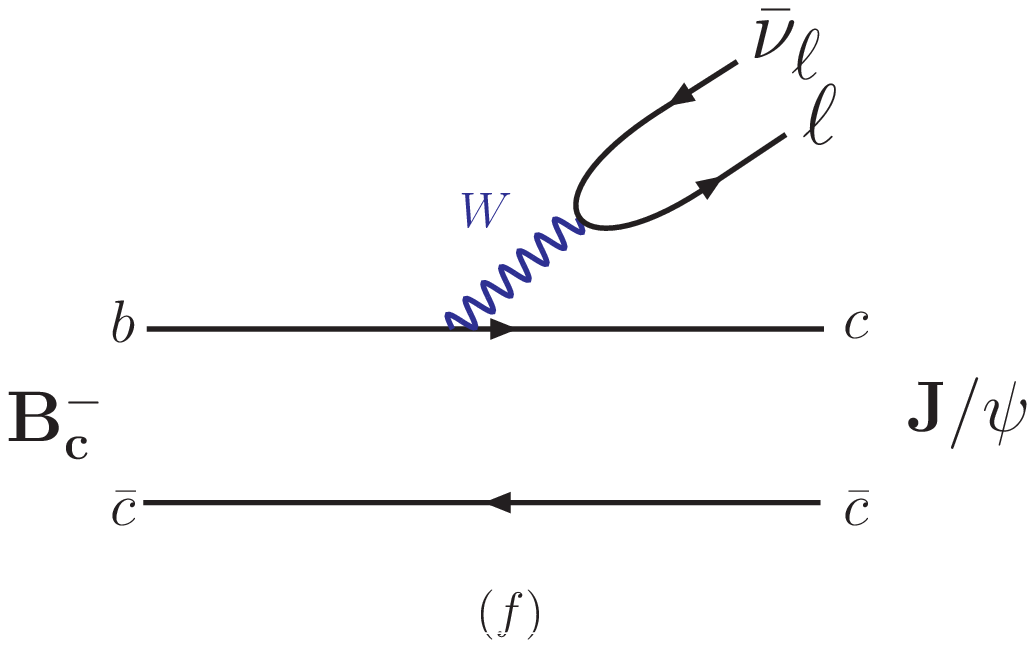}
\caption{Diagrams for the $B$ and $B_c$ decays with the formations of the $c\bar c$ pair, where
$(a)$, $(b)$ and $(c)$ correspond to the $B\to X_c^0 M$, $B^-_c\to X_c^0 M$,
and $B^-_c\to X_c^0 \ell \bar \nu_\ell$ decays,
while $(d)$, $(e)$ and $(f)$ the $B\to J/\psi M$, $B^-_c\to J/\psi M$,
and $B^-_c\to J/\psi \ell \bar \nu_\ell$ decays, respectively.}\label{fig1}
\end{figure}
In terms of the effective Hamiltonians at quark level for the $b\to c\bar c q$,
$b\to c \bar u q$, and $b\to c \ell\bar \nu_\ell$ transitions in Fig.~\ref{fig1},
the amplitudes of  the $B_c^-\to M_c M$, $B\to M_c M$,
and $B_c^-\to M_c \ell^-\bar \nu_\ell$ decays can be factorized as~\cite{ali}
\begin{eqnarray}\label{amp1}
{\cal A}(B_c^-\to M_c M)&=&
i\frac{G_F}{\sqrt 2}V_{cb}V_{uq}^*a_1f_M
\langle M_c|\bar c\, \slashed q(1-\gamma_5) b|B_c^-\rangle\,,\nonumber\\
{\cal A}(B\to M_c M)&=&
\frac{G_F}{\sqrt 2}V_{cb}V_{cq}^* a_2 m_{M_c}f_{M_c}
\langle M|\bar q\slashed \varepsilon(1-\gamma_5) b|B\rangle\,,\nonumber\\
{\cal A}(B_c^-\to M_c \ell^-\bar \nu_\ell)&=&
\frac{G_F V_{cb}}{\sqrt 2}
\langle M_c|\bar c \gamma_\mu(1-\gamma_5) b|B_c^-\rangle
\bar \ell\gamma^\mu(1-\gamma_5)\nu_\ell\,,
\end{eqnarray}
respectively,
where $\slashed q=q^\mu \gamma_\mu$,
$\slashed \varepsilon=\varepsilon^{\mu *}\gamma_\mu$,
$q=d(s)$ for $M=\pi^-(K^-)$, $M_c=(J/\psi, X_c^0)$, $\ell=(e^-,\mu^-,\tau^-)$,
$G_F$ is the Fermi constant, $V_{ij}$ are the CKM matrix elements, and
$a_{1,2}$ are the parameters to be determined.
In Eq.~(\ref{amp1}),  the decay constant, four momentum vector, and four polarization
$(f_{M_{(c)}},q^\mu,\varepsilon^{\mu *})$ are defined by
\begin{eqnarray}\label{fXc}
\langle M|\bar q\gamma_\mu\gamma_5 u|0\rangle&=&-if_M q^\mu\,,\nonumber\\
\langle J/\psi|\bar c\gamma_\mu c|0\rangle&=&m_{J/\psi}f_{J/\psi}\varepsilon_\mu^*\,,\nonumber\\
\langle X_c^0|\bar c\gamma_\mu\gamma_5 c|0\rangle&=&m_{X_c^0}f_{X_c^0}\varepsilon_\mu^*\,,
\end{eqnarray}
while the matrix elements of
the $B\to (M,J/\psi,X_c^0)$ transitions can be parametrized as~\cite{Wang:2007sxa}
\begin{eqnarray}\label{ff1}
\langle M| \bar q \gamma^\mu b|B\rangle&=&\bigg[(p_B+p_M)^\mu-\frac{m^2_B-m^2_M}{t}q^\mu\bigg]
F_1^{BM}(t)+\frac{m^2_B-m^2_M}{t}q^\mu F_0^{BM}(t)\,,\nonumber\\
\langle J/\psi|\bar c\gamma_\mu b|B_c^-\rangle&=&\epsilon_{\mu\nu\alpha\beta}
\varepsilon^{\ast\nu}p_{B_c}^{\alpha}p_{J/\psi}^{\beta}\frac{2V(t)}{m_{B_c}+m_{J/\psi}}\;,\nonumber\\
\langle J/\psi|\bar c\gamma_\mu \gamma_5 b|B_c^-\rangle
&=&i\bigg[\varepsilon^\ast_\mu-\frac{\varepsilon^\ast\cdot q}{t}q_\mu\bigg](m_{B_c}+m_{J/\psi})A_1(t)
 + i\frac{\varepsilon^\ast\cdot q}{t}q_\mu(2m_{J/\psi})A_0(t)\nonumber\\
&-&i\bigg[(p_{B_c}+p_{J/\psi})_\mu-\frac{m^2_{B_c}
-m^2_{J/\psi}}{t}q_\mu \bigg](\varepsilon^\ast\cdot q)\frac{A_2(t)}{m_B+m_{J/\psi}}\;,\nonumber\\
\langle X_c^0|\bar c\gamma_\mu \gamma_5 b|B_c^-\rangle&=&-\epsilon_{\mu\nu\alpha\beta}
\varepsilon^{\ast\nu}p_{B_c}^{\alpha}p_{X_c^0}^{\beta}\frac{2iA(t)}{m_{B_c}-m_{X_c^0}}\;,\nonumber\\
\langle X_c^0|\bar c\gamma_\mu b|B_c^-\rangle
&=&-\bigg[\varepsilon^\ast_\mu-\frac{\varepsilon^\ast\cdot q}{t}q_\mu\bigg](m_{B_c}-m_{X_c^0})V_1(t)
 - \frac{\varepsilon^\ast\cdot q}{t}q_\mu(2m_{X_c^0})V_0(t)\nonumber\\
&+&\bigg[(p_{B_c}+p_{X_c^0})_\mu-\frac{m^2_{B_c}
-m^2_{X_c^0}}{t}q_\mu \bigg](\varepsilon^\ast\cdot q)\frac{V_2(t)}{m_B-m_{X_c^0}}\;,
\end{eqnarray}
respectively,
where $q=p_B-p_{M_{(c)}}$, $t\equiv q^2$, and
 $(F_{1,2}, A_{(i)},V_{(i)})$ with $i=0,1,2$ are the form factors.

\section{Numerical Results and Discussions}
In our numerical analysis, we use the Wolfenstein parameterization
for the CKM matrix elements in Eq.~(\ref{amp1}), given by
$V_{cb}=A\lambda^2$, $V_{ud}=V_{cs}=1-\lambda^2/2$, and $V_{us}=-V_{cd}=\lambda$,
with~\cite{pdg}
\begin{eqnarray}
(\lambda,\,A,\,\rho,\,\eta)=(0.225,\,0.814,\,0.120\pm 0.022,\,0.362\pm 0.013)\,.
\end{eqnarray}
The parameters $a_{1,2}$, decay constants and form factors,
adopted from Refs.~\cite{Geng:2016drz,Hsiao:2016vck},
\cite{pdg,Becirevic:2013bsa}, and \cite{MS,Wang:2007sxa} are as follows:
\begin{eqnarray}
&&(a_1,a_2)=(1.05^{+0.12}_{-0.06},0.268\pm 0.004)\,,\nonumber\\
&&(f_\pi,f_K,f_{J/\psi})=(130.4\pm 0.2,156.2\pm 0.7,418\pm 9)\;\text{MeV}\,,\nonumber\\
&&(F_1^{B\pi}(0),F_1^{BK}(0),F_1^{BsK}(0))=(0.29,0.36,0.31)\,,
\end{eqnarray}
where the form factors correspond to the reduced matrix elements
derived from Eqs.~(\ref{amp1}) and (\ref{ff1}), given by
\begin{eqnarray}\label{Ff_Xc}
\langle M|\bar q\slashed \varepsilon b|B\rangle&=&
\varepsilon\cdot (p_B+p_M)F_1^{BM}\,. 
\end{eqnarray}
The momentum dependence for $F_{1}^{BM}(q^2)$ from Ref.~\cite{MS} is taken as
\begin{eqnarray}\label{form2}
F^{BM}_1(t)=
\frac{F^{BM}_1(0)}{(1-\frac{t}{M_V^2})(1-\frac{\sigma_{11} t}{M_V^2}+\frac{\sigma_{12} t^2}{M_V^4})}\,,\;
\end{eqnarray}
with $\sigma_{11}=(0.48,0.43,0.63)$, $\sigma_{12}=(0,0,0.33)$
and $M_V=(5.32,5.42,5.32)$ GeV for $B\to \pi$, $B\to K$ and $\bar B^0_s\to K$, respectively.
With ${\cal B}(B^-\to X_c^0 K^-)/{\cal B}(B^-\to J/\psi K^-)=0.22\pm 0.09$ from Eq.~(\ref{data1}),
we obtain $f_{X_c^0}=(234\pm 52)$~MeV, which is lower than
$f_{X_c^0}=(335,329^{+111}_{-\;\;95})$~MeV~\cite{Liu:2007uj,Wang:2007sxa}
from perturbative and light-front QCD models, respectively.
The momentum dependences for the $B_c\to M_c$ transition form factors
are given by~\cite{Wang:2008xt}
\begin{eqnarray}\label{F_Mc}
f(t)=f(0)\text{exp}(\sigma_1 t/m_{B_c}^2+\sigma_2 t^2/m_{B_c}^4)\,,
\end{eqnarray}
where the values of $f(0)=(V_{(i)}(0),A_{(i)}(0))$ and $\sigma_{1,2}$ in Table~\ref{ff0}
are from Refs.~\cite{Wang:2007sxa} and \cite{Wang:2008xt}, respectively.
%
\begin{table}[t]
\caption{The $B_c\to (J/\psi,X_c^0)$ form factors at $t=0$ and
$\sigma_{1,2}$ for the momentum dependences in Eq.~(\ref{F_Mc}).
}\label{ff0}
\begin{tabular}{|c||c|ccc|}
\hline
$B_c\to (J/\psi,X_c^0)$&$f(0)$~\cite{Wang:2007sxa}
&$\sigma_1$&$\sigma_2$&\cite{Wang:2008xt}\\\hline
$(V,A)$         &$(0.87\pm 0.02,0.36\pm 0.04)$&2.46&0.56&\\  
$(A_0,V_0)$ &$(0.57\pm 0.02,0.18\pm 0.03)$&2.39&0.50&\\  
$(A_1,V_1)$ &$(0.55\pm 0.03,1.15\pm 0.07)$&1.73&0.33&\\  
$(A_2,V_2)$ &$(0.51\pm 0.04,0.13\pm 0.02)$&2.22&0.45&\\
\hline
\end{tabular}
\end{table}
\begin{table}[b]
\caption{The branching ratios of the $B_{c}\to J/\psi (M,\ell\bar \nu_\ell)$ decays,
where the first (second) errors of our results are from the form factors ($a_1$).} \label{tabJ}
\begin{tabular}{|c|cc|}
\hline
decay modes  &  our results & QCD models\\\hline
$B_c^-\to J/\psi \pi^-$
&$(10.9\pm 0.8^{+2.6}_{-1.2})\times 10^{-4}$
& $(20^{+8+0+0}_{-7-1-0})\times 10^{-4}$~\cite{Wang:2007sxa}\\
$B_c^-\to J/\psi K^-$
&$(8.8\pm 0.6^{+2.1}_{-1.0})\times 10^{-5}$
& $(16^{+6+0+0}_{-6-1-0})\times 10^{-5}$~\cite{Wang:2007sxa}\\
$B_c^-\to J/\psi e^-\bar \nu_e$
&$(1.94\pm 0.20)\times 10^{-2}$
& $(1.49^{+0.01+0.15+0.23}_{-0.03-0.14-0.23})\times 10^{-2}$~\cite{Wang:2008xt}\\
$B_c^-\to J/\psi\mu^-\bar \nu_\mu$
&$(1.94\pm 0.20)\times 10^{-2}$
& $(1.49^{+0.01+0.15+0.23}_{-0.03-0.14-0.23})\times 10^{-2}$~\cite{Wang:2008xt}\\
$B_c^-\to J/\psi\tau^-\bar \nu_\tau$
&$(4.47\pm 0.48)\times 10^{-3}$
& $(3.70^{+0.02+0.42+0.56}_{-0.05-0.38-0.56})\times 10^{-3}$~\cite{Wang:2008xt}\\
\hline
\end{tabular}
\end{table}
Our results for the branching ratios of $B_c^-\to J/\psi (\pi^-,K^-,\ell^-\bar \nu_\ell)$
are shown in Table~\ref{tabJ}.

 From Table~\ref{tabJ}, we see that our numerical values of ${\cal B}(B_c^-\to J/\psi \pi^-)$ and
${\cal B}(B_c^-\to J/\psi K^-)$ are about a factor 2 smaller than those in Ref.~\cite{Wang:2007sxa},
where the calculations were done only by the leading-order contributions in the $1/{m_{B_c}}$ 
expansion~\footnote{We thank  the authors in Ref.~\cite{Wang:2007sxa} for the useful communication.}.
From the table, we get that
${\cal B}(B_c^-\to J/\psi \pi^-)/{\cal B}(B_c^-\to J/\psi K^-)=0.078\pm 0.027$,
which agrees with ${\cal R}_{K/\pi}$ in Eq.~(\ref{data2}),
demonstrating the validity of the factorization approach.
By taking ${\cal B}(B_c^-\to J/\psi \pi^-)$ as the theoretical input in Eq.~(\ref{data2}),
we extract that
\begin{eqnarray}\label{semi}
&&f_c/f_u=(6.4\pm 1.9)\times 10^{-3}\,,
\end{eqnarray}
which can be useful to determine the experimental data,
such as that in Eq.~(\ref{data2}).
\begin{table}[b]
\caption{The branching ratios for the $B_{(c)}\to X_c^0 M$ and 
$B_c\to X_c^0\ell\bar \nu_\ell$ decays. For our results,
the first errors come from $(f_{X_c^0},f(0))$, while the second ones
from $(a_1,a_2)$.} \label{tab1}
\begin{tabular}{|c|cc|}
\hline
decay modes  &  our results & QCD models\\\hline
$B^-\to X^0_c \pi^-$
&$(11.5^{+5.7}_{-4.5}\pm 0.3)\times 10^{-6}$ & -----\\
$B^-\to X^0_c K^-$
&$(2.3^{+1.1}_{-0.9}\pm 0.1)\times 10^{-4}$ & $(7.88^{+4.87}_{-3.76})\times 10^{-4}$~\cite{Liu:2007uj} \\
$\bar B^0\to X^0_c \pi^0$
&$(5.3^{+2.6}_{-2.1}\pm 0.2)\times 10^{-6}$ & -----\\
$\bar B^0\to X^0_c \bar K^0$
&$(2.1^{+1.0}_{-0.8}\pm 0.1)\times 10^{-4}$ & -----\\
$\bar B^0_s\to X^0_c \bar K^0$
&$(11.4^{+5.6}_{-4.5}\pm 0.3)\times 10^{-6}$& -----\\
$B_c^-\to X^0_c \pi^-$
&$(6.0^{+2.2+1.4}_{-1.8-0.7})\times 10^{-5}$
& $(1.7^{+0.7+0.1+0.4}_{-0.6-0.2-0.4})\times 10^{-4}$~\cite{Wang:2007sxa}\\
$B_c^-\to X^0_c K^-$
&$(4.7^{+1.7+1.1}_{-1.4-0.5})\times 10^{-6}$
& $(1.3^{+0.5+0.1+0.3}_{-0.5-0.2-0.3})\times 10^{-5}$~\cite{Wang:2007sxa}\\
$B_c^-\to X_c^0 e^-\bar \nu_e$
&$(1.35\pm 0.18)\times 10^{-3}$
& $(6.7^{+0.9+0.0+0.1+0.5+2.3+0.7}_{-0.5-0.0-0.0-0.5-2.6-0.7})\times 10^{-3}$~\cite{Wang:2007fs}\\
$B_c^-\to X_c^0 \mu^-\bar \nu_\mu$
&$(1.35\pm 0.18)\times 10^{-3}$
& -----\\
$B_c^-\to X_c^0\tau^-\bar \nu_\tau$
&$(6.5\pm 0.9)\times 10^{-5}$
& $(3.2^{+0.5+0.0+0.0+0.2+1.1+0.4}_{-0.2-0.2-0.0-0.2-1.3-0.3})\times 10^{-4}$~\cite{Wang:2007fs}\\
\hline
\end{tabular}
\end{table}

For the $B\to X_c^0 (\pi,K)$ decays, the results are given in Table~\ref{tab1}.
While $f_{X_c^0}=(234\pm 52)$~MeV leads to
${\cal B}(B^-\to X^0_c K^-)=(2.3^{+1.1}_{-0.9}\pm 0.1)\times 10^{-4}$
in accordance with the data, we predict that
${\cal B}(B^-\to X_c^0\pi^-)=(11.5\pm 5.7)\times 10^{-6}$,
${\cal B}(\bar B^0\to X_c^0\bar K^0)=(2.1\pm 1.0)\times 10^{-4}$, and
${\cal B}(\bar B^0_s\to X_c^0\bar K^0)=(11.4\pm 5.6)\times 10^{-6}$,
which are accessible to the experiments at the LHCb.
Besides, our results of
${\cal B}(\bar B^0_s\to X^0_c \bar K^0)\simeq {\cal B}(B^-\to X^0_c \pi^-)$ and
${\cal B}(\bar B^0\to X^0_c \pi^0)\simeq {\cal B}(B^-\to X^0_c \pi^-)/2$
in Table~\ref{tab1} are also supported by the $SU(3)$ and isospin symmetries, respectively.
With the form factors adopted from Ref.~\cite{Wang:2007sxa}, we calculate that
${\cal B}(B_c^-\to X^0_c \pi^-)=(6.0\pm 2.6)\times 10^{-5}$ and
${\cal B}(B_c^-\to X^0_c K^-)=(4.7\pm 2.0)\times 10^{-6}$, which
are 2-3 times smaller than the results from the same reference.
The differences are again reconciled after keeping 
 the next-leading order   contributions in the $1/{m_{B_c}}$ expansion.

\begin{figure}[t]
\centering
\includegraphics[width=3in]{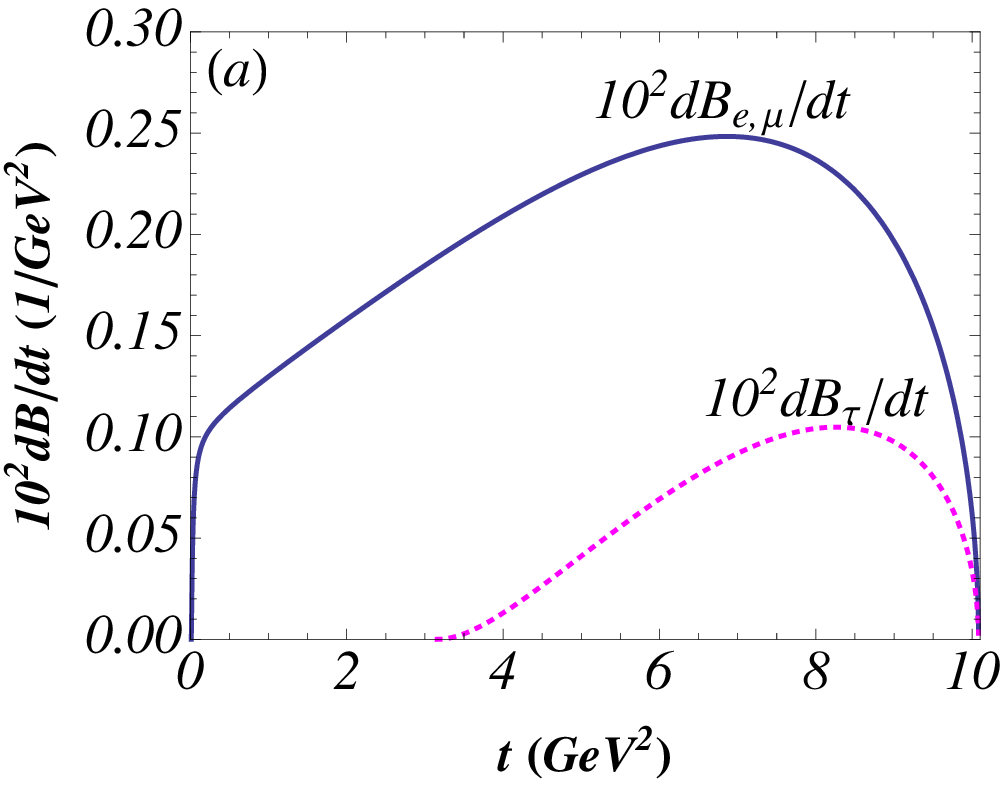}
\includegraphics[width=3in]{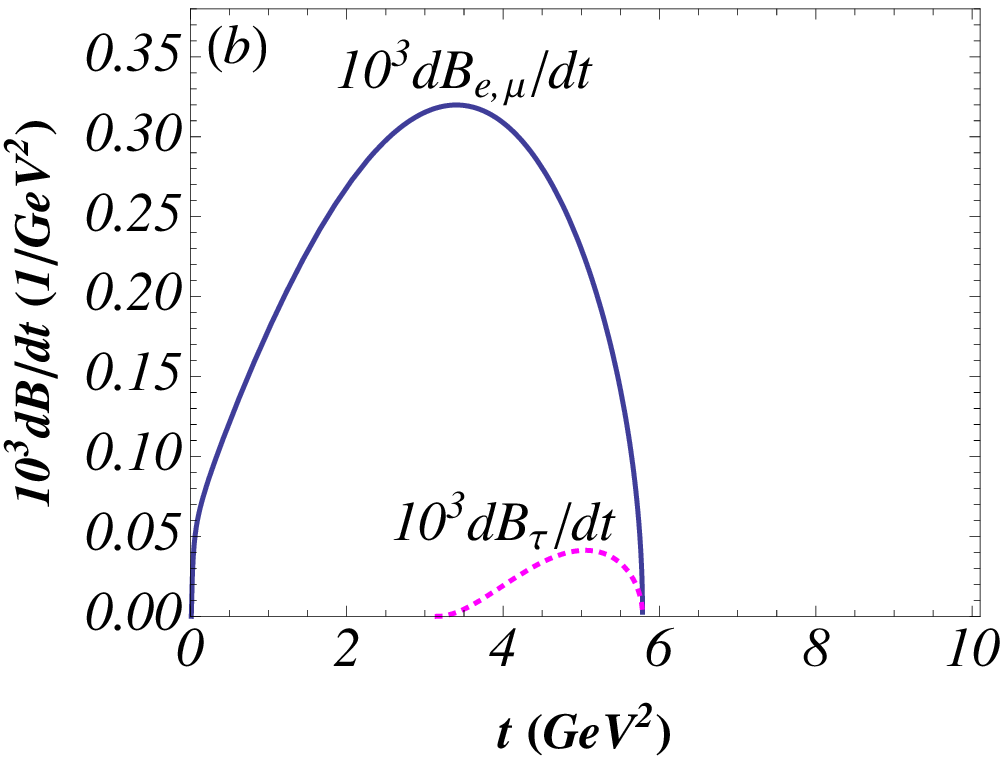}
\caption{The spectra of
the semileptonic 
(a) $B_c^-\to J/\psi\ell^-\bar \nu_\ell$ and 
(b) $B_c^-\to X_c^0\ell^-\bar \nu_\ell$ decays,
where the solid and dotted lines correspond to 
$\ell=(e,\mu)$ and $\ell=\tau$, respectively.}\label{spec}
\end{figure}

For the semileptonic $B_c^-\to M_c \ell^- \bar \nu_\ell$ decays,
${\cal B}(B_c^-\to J/\psi e\bar \nu_e)={\cal B}(B_c^-\to J/\psi \mu\bar \nu_\mu)
=(1.94\pm 0.20)\times 10^{-2}$
is due to the both negligible electron and muon masses, 
of which the numerical value is close to those from 
Refs.~\cite{Wang:2008xt,Huang:2007kb}. Note that
by taking ${\cal B}(B_c^-\to J/\psi \pi^-)$ as the theoretical input in Eq.~(\ref{data2}),
we derive that
\begin{eqnarray}
{\cal B}(B_c^-\to J/\psi \mu^-\bar \nu_\mu)=(2.3\pm0.6)\times 10^{-2}\,,
\end{eqnarray}
which agrees with the above theoretical prediction. 
For the $\tau$ mode, which suppresses the phase space due to the heavy $m_\tau$,
we obtain ${\cal B}(B_c^-\to J/\psi \tau^-\bar \nu_\tau)=(4.47\pm 0.48)\times 10^{-3}$.
The ratio of 
${\cal B}(B_c^-\to X_c^0 e^-\bar \nu_e)/
{\cal B}(B_c^-\to X_c^0\tau^-\bar \nu_\tau)\simeq 1/20$
is close to that in Ref.~\cite{Wang:2007fs}, but 
${\cal B}(B_c^-\to X_c^0 e^-\bar \nu_e)=(1.35\pm 0.18)\times 10^{-3}$
is apparently 4-5 times smaller than that in Ref.~\cite{Wang:2007fs},
though with uncertainties the two results overlap with each other.
With the spectra of $B_c^-\to (J/\psi,X_c^0)\ell^-\bar \nu_\ell$  in Fig.~\ref{spec},
our results can be compared to the recent studies on the semileptonic
$B_c$ cases in Refs.~\cite{Wang:2015rcz,semi2} for the $XYZ$ states.
%

\section{Conclusions}
In sum, we have studied the $B_{(c)}\to M_c(\pi,K)$ and
$B_{c}\to M_c\ell^-\bar \nu_\ell$ decays with $M_c=J/\psi$ and $X_c^0\equiv X^0(3872)$.
We have presented that
${\cal B}(B^-\to X_c^0\pi^-,X_c^0 K^-)=(11.5\pm 5.7)\times 10^{-6}$
and $(2.3\pm 1.1)\times 10^{-4}$, and
${\cal B}(B_c^-\to X_c^0\pi^-,X_c^0 K^-)=(6.0\pm 2.6)\times 10^{-5}$
and $(4.7\pm 2.0)\times 10^{-6}$. With ${\cal B}(B_c^-\to J/\psi \pi^-)=(10.9\pm 2.6)\times 10^{-4}$
as the theoretical input, the extractions from the data have shown that 
$f_c/f_u=(6.4\pm 1.9)\times 10^{-3}$ and
${\cal B}(B_c^-\to J/\psi \mu^-\bar \nu_\mu)=(2.3\pm0.6)\times 10^{-2}$.
We have estimated 
${\cal B}(B_c^-\to X_c^0 \ell^-\bar \nu_\ell)$ with $\ell=(e^-,\mu^-,\tau^-)$
to be $(1.35\pm 0.18)\times 10^{-3}$, $(1.35\pm 0.18)\times 10^{-3}$, 
and $(6.5\pm 0.9)\times 10^{-5}$, respectively.

\section*{ACKNOWLEDGMENTS}
The work was supported in part by
National Center for Theoretical Sciences,
National Science Council (NSC-101-2112-M-007-006-MY3), and
MoST (MoST-104-2112-M-007-003-MY3).

\end{document}